# Absolute Prioritization of Planetary Protection, Safety, and Avoiding Imperialism in All Future Science Missions: A Policy Perspective

Primary Author: Monica Vidaurri (George Mason University, NASA GSFC)
Contact: monica.r.vidaurri@nasa.gov

Co-Authors:
Alia Wofford (Howard University, NASA GSFC)
Jonathan Brande (University of Maryland, NASA GSFC)
Gabriel Black-Planas (JMHS)
Shawn Domagal-Goldman (NASA GSFC)
Jacob Haqq-Misra (Blue Marble Space Institute of Science)

Co-Signers:
Peter Plavchan (George Mason University)
Danny LeBrun (Lockheed Martin)
Chuanfei Dong (Princeton University)
Daniel Angerhausen (Bern University)
Sarah Tuttle (University of Washington, Seattle)
Kathryn Denning (York University)
Sofia Sheikh (Penn State University)
Daria Pidhorodetska (NASA GSFC)
Giada Arney (NASA GSFC)
Amy C. Barr (Planetary Science Institute)
Sam Ragland (W.M. Keck Observatory)
John Debes (STScI)
Lia Corrales (University of Michigan)
Aparna Venkatesan (University of San Francisco)
Meredith Durbin (University of Washington, Seattle)
Joleen Carlberg (STScI)
Robert Simon (Kennesaw State University)
Sandra Bastelberger (University of Maryland)
Erin Maier (Steward Observatory, University of Arizona)
D. J. Teal (University of Maryland)
Gourav Khullar (University of Chicago)
R. Deno Stelter (University of California, Santa Cruz)

## 1. Abstract.

The prioritization and improvement of ethics, planetary protection, and safety standards in the astro-sciences is the most critical priority as our scientific and exploratory capabilities progress, both within government agencies and the private sector. These priorities lie in the belief that every single science mission – crewed or non-crewed, ground-based or not – should heed strict ethical and safety standards starting at the very beginning of a mission. Given the inevitability of the private sector in influencing future crewed missions both in and beyond low-Earth orbit, it is essential to the science community to agree on universal standards of safety, mission assurance, planetary protection, and especially anti-colonization. These issues will impact all areas of space science. Examples that are particularly relevant to the Astro2020 Decadal Survey include but are not limited to: light pollution from satellites, the voices and rights of Native people when constructing telescopes on their lands, and the need to be cognizant of contamination when searching for and exploring habitable environments beyond Earth. The existence of oversight bodies to enforce planetary protection and communication between public, private, and academia is necessary for this proposal. Delegation of power and strict communication standards not only to protect the lives of the explorers, but protect the environments of wherever humanity decides to venture. Opening up the multidisciplinary approach of space exploration to international law and governance regarding planetary protection, safety, mission assurance, and creating comprehensive and ethical standards across all space faring institutions is needed for the future of space exploration. Agreement and enforcement by the United Nations Office of Outer Space Affairs (UNOOSA) and the cooperation of participating governments will also prove critical in regulating and improving standards for future science missions. Ultimately, moving international space law and domestic space policy from a reactive nature to a proactive one will ensure the future of space exploration is one that is safe, transparent, and anti-imperialist. The prioritization of safety, planetary protection, and ethical practices of space exploration and its subsets is heavily dependent on a clear, progressive, and precautionary approach to international and domestic space law.

## 2. Motivation.

While planetary protection standards are mandatory for all exploration missions, it is rare that scientists, scholars, and scientific policy writers cite or actively integrate these standards in mission conception and execution. NASA has protectionary standards set in place for robotic missions, but a 2018 NAS report describes these standards as "inadequate".[1] For crewed missions, standards are non-existent and will be the most difficult to create and implement due to the human body's physiological limitations. As stated in the NAS Review of Planetary Protection Strategies, the progression of these standards has not gained much ground since the days of the Viking landers.[1] Additionally, the new and quickly evolving private industry and non-federal space actors largely experience regulatory gaps when it comes to protection standards, adherence to international space customs, and subsequent bioethical standards.

With Mars2020, Europa Clipper, and the inevitable exploration of other bodies in the solar system such as ocean worlds, regulatory entities such as NASA, UN committees (COSPAR and UNOOSA), and all other international partners must lead the effort to create and enforce planetary protection standards, cultivate a pathway with delegated authority for crewed and uncrewed space exploration campaigns, including tourism and commercial activities. In addition to setting international norms to be used in space law, all leaders and participants in space exploration must also adopt anti-colonization standards and protocol to ensure equal and fair participation in space. This will allow for only purely peaceful scientific purposes for exploration while ensuring minimal contamination.

Given the importance of planetary protection and new standards of safety with respect to technological and scientific advancement, it is vital to the international space community to work to improve international law and space policy. By actively implementing strict safety customs from the individual level, the proper chain of command via the respective government can ensure enforcement of standards in the form of an independent and strengthened planetary protection office (PPO). In other words, a PPO would act as an advisory and authoritative body. In conjunction, collaboration between PPOs and the astro-sciences is equally important in improving microbial detection and decontamination technologies.

### 3. Strengthening the Framework – Areas of Consideration

Universalization of planetary protection, safety, and ethical standards is crucial for all future space exploration directives and missions. In this case, universalization of standards is heavily reliant upon building and maintaining a regulatory and communicative global infrastructure. Delegated roles are assigned vertically to provide adequate checks and continuous risk assessment starting after mission conception and throughout the lifetime of the mission. Implementing this infrastructure would bridge the regulatory gap between private and public sectors. Delegation of safety procedures and anti-colonization would be best executed by a PPO, ethics committees, active ad-hoc committees, and other representatives involved in mission communication and enforcement. Without the implementation of practicing norms and customs from mission management, any formal establishment of law, from the agency level to international level, will not have a solid practical foundation. Consequently, laws will be established with no prior background of how these norms can be improved and how they help.

#### Bridging the Regulatory Gap and the Role of the Space Agency

The Outer Space Treaty (OST) places the responsibility of understanding and adhering to exploration safety and bioethical standards for non-government entities on the respective government(s) and relative agencies representing the entity.[3,4] For example, to establish the norm of proactive safety and ethics standards, NASA can take it upon themselves to outline and communicate with Congress its full planetary protection and bioethics strategies at the approval of every science mission. Meanwhile, strategies continue to remain compliant with UNOOSA standards including astronaut safety, contamination, non-militarization and anti-colonization, and transparency of technology. Therefore, actions must be taken by governments ultimately responsible for these exploratory actions to allow the authority of the agency to constantly monitor, communicate, and enforce safety and protectionary regulations including private sector partnerships. These actions will effectively ensure enforcement of standards in addition to those stated in NASA Procedural Requirements (NPR) 8020.12D, outlining the use of NASA (agency) funding from non-NASA entities only at the demonstration of adherence to policies regarding planetary protection as set forth by COSPAR.[4]

Section 2.2.3. of NPR8020.12D requires non-NASA entities to submit future protectionary procedural outlines to the NASA Planetary Protection Office. In the proposed framework, submissions are approved by the agency and the PPO, and enforced accordingly. Protectionary procedure reports would be submitted to the PPO and must include planetary protection procedures outlined by the agency and highlight any extra items set forth by the non-agency entity. For this reason, it would benefit every agency to have a communicative and transparent PPO (a function of the agency) or equivalent entity to follow and modify these procedures as needed over the course of a mission, while remaining compliant with and communicating the concerns of ethical committees and liaisons.

Furthermore, Section 2.4 of NPR8020.12D states that, at the request of the PPO, mission management shall make arrangements to allow PPO representatives to be present during transport, decontamination, hardware and environment assessments, and writing of documentation relative to safety and protection.[4] To strengthen this with regard to the proposed framework, the PPO and any relevant ethics and safety committees should be present from conception (which includes mission categorization) through mission completion, with the PPO scheduling regular assessments of decontamination procedure, adherence to standards, hardware, documentation, and any other mission detail falling under the jurisdiction of the PPO. Communication and enforcement/improvement of standards can be achieved with ad-hoc committees or representatives established within a mission team as a direct line to a PPO for the purposes of upholding strict protectionary standards and norms.

### Reactive to Proactive Safety Procedures and their Jurisdiction

With the establishment of PPO as an authoritative body, the proposed standards can be accomplished. In conjunction with a PPO, any existing ethical and safety committees or liaisons sharing this authority and necessary responsibilities throughout the mission may take more protective roles. This includes advising and assessing implications a mission poses to the future of planetary protection, ethics, and safety, while reporting these implications to the representative government and UNOOSA. The applicability, combined with language changes enforcing the role of the PPO and relevant committees/liaisons as objective and mandatory in all exploration missions – agency/government and non-agency – proves the reinstatement of NPR8020.12D after expiration as procedural standard and international custom would benefit the progression of these standards. Moving forward, reinstatement of this procedural requirement, or the creation of a new requirement including language applicable to crewed and LEO missions, is needed.

NASA Policy Directive (NPD) 8020.7G, accounting for the contamination control of both outbound and inbound spacecraft, would also prove adequate procedural foundation on an international level.[2] To illustrate, the directive must adopt executable language and delegation of authority similar to that laid out in the previous section upon a renewal, but should ultimately prove useful for law language. Reinstatement to include quarantine and contamination control (along with the necessary research for this) for crewed missions would also prove useful in this sense as the space science community progresses further into human exploration.

Furthermore, having an inter-agency framework with respect to the safety and biothreat control of spacefaring missions can prove to be another method of obtaining and maintaining a proactive safety policy, as opposed to the current reactive system. For example, in the United States, partnership with the Department of Energy, the National Institutes of Health, and relevant agencies on comprehensive study and assessment of microbiome survivability in closed systems such as the ISS not only helps relieve NASA of studying these effects alone, but helps build an intellectual framework that is transparent and sharing in nature that all parties can benefit from, and that can be used for other matters, such as the United States National Biodefense Strategy, or future space tourism safety procedures.

### Planetary Protection and Interstellar Communication

An extension of the planetary protection framework could also apply to the consideration of sending messages that are intended to be received by other civilizations (often referred to as Messages to Extraterrestrial Intelligent life, or METI). The discovery of exoplanets with orbital and size/mass properties that could allow globally habitable conditions, and plans being submitted to Astro2020 to construct telescopes that could confirm the habitability of such worlds and search them for signs of life, increase the relevance of METI. The age of exoplanets has further motivated renewed efforts at the Search for Extraterrestrial Intelligence (or SETI) by searching stellar systems for anomalous radio or optical signals. Some organizations also explicitly advocate for METI as an effort to increase the likelihood of discovering extraterrestrial intelligent life. Some scholars have voiced opposition to any attempts at METI by citing ethical concerns, such as the possible negative consequences of initiating such contact or the inability to readily decide a spokesperson or message representative of Earth as a whole (see, e.g. "The METI Debate" convened in 2010 by the Royal Society)[5]. Others have pointed out that low-cost and low-power METI today could provide beneficial preparation for the future, while any actual contact with extraterrestrials could also be immensely positive or wholly neutral, in addition to negative[6].

Past METI efforts and discussions have also expressed the potential positive benefits from constructing a message that represents humanity. They acknowledge that any message sent by Earth should be representative of the diversity of life on Earth. As a result, even research into what kinds of messages would be sent can be diverse, inclusive, and multicultural. Such efforts - if connected to broader topics and related careers - could lead to more inclusivity and diversity in exoplanets and astrophysics. However, any benefits from such an effort will only be realized if the message construction is intrinsically inclusive. Further, the daunting ethical issue remains of how to weigh the multitude of voices on Earth in the context of socioeconomic inequalities.

All of this demands ethical and moral infrastructure, and that they be implemented as a fundamental part of any SETI or METI project.

### Anti-Colonization

"Seeking to prevent a new form of colonial competition" was the very premise to the OST.[3] As such, the creation of a norm and/or resolution barring non-scientific human settlement on any celestial body is encouraged to be established and adopted as custom. It is critical that international law adopts a custom of good faith in anti-colonization, similar to militarization, with the presentation of these customs from a majority of each state and agency participating in space exploration, as well as the establishment of explicit anti-colonialism as a clause of the OST.[7] To achieve this, it is important for space-faring nations to remember that colonization is a structure; an institutionalized virtue stemming from exploitation of (often native) populations to preserve a central selfish ideal of preserving the way of life as seen by the colonizing nation. In addition, states party to and responsible for elements of scientific stations should uphold respective jurisdiction over their contributions, utilisation, and liability rights of facilities, equipment, technologies, personnel, and territories.

It is paramount that the astro community not only recognize colonialism and imperialism in the premise of settling other planets/celestial bodies and their orbits, but also work together to prevent further colonialism on Earth. A proper check to colonialism from the astro community takes the example of the Thirty Meter Telescope (TMT), as well as the other telescopes that have been built on Native lands in Mauna Kea, Hawaii.[8] Management of the current set of telescopes in Mauna Kea have established the norms of paying rent, creating a work pipeline for the Native people, creating education outreach programs for Native youth, and actively facilitating communication with Native tribes and businesses, thus gaining the trust and approval of the Native people to build all current telescope operations. However, the continued controversy around Mauna Kea and the TMT should serve as a new regulatory area of consideration: one that includes "no further development, creat[ing] a community-based management authority including environmentalists, native Hawaiians, regulatory agencies, and collect[ing] rent."[9] Ground-based observing, as well as other astro-science related developments, cannot and should not take precedence over lands belonging to Native populations, or areas where telescope development is potentially harmful towards native ecosystems and the environment.

Thus, it is imperative that anti-imperialist standards and methods of thinking begin in the early stages of mission concepcion, along with safety and ethics, and from the PI/individual level through the agency/global level, which can be facilitated and held accountable by the creation/utilization of ethics committees and liaisons.

In addition, anti-militarization articles and clauses of space allow for the existence of war facilitators[+] in space, but no scientific justification can be made for the operation of such war facilitators for peaceful purposes on other celestial bodies. Therefore, anti-colonization is to include anti-establishment of all military-like installments on celestial objects, including those with potential

---

[+] War facilitators are defined for these purposes as inherently peaceful objects, personnel, and low-earth orbiting craft utilizing terra nullius of outer space and airspace, that have multi-capability instrumentation and purpose, including but not limited to: global positioning systems (GPS), reconnaissance drones, astronaut survival weapons, and communications satellites with multi-level access.

war facilitators. This clause shall exclude mining equipment put forth and approved by a PPO and UNOOSA, along with specific uses that a mission must adhere to where the use of mining and other potential war-facilitating science equipment for purposes other than those stated in the mission is met with discipline as understood by an international space community.

## 4. Ensuring the Framework – Norms and Strategies

Lobbying

In order to preserve and uphold a more future-oriented, safe, and comprehensive interactive legal framework between the public and private sectors and international partners, it would benefit governments to not allow lobbying efforts to streamline certifications, permits, and renewals of legal documentation allowing access to LEO and other celestial bodies/orbits to pass. Though transparency of technologies used by the private sector is critical, it is not needed with the correct framework and agency technology, planetary protection, and ethics officers maintaining constant communication and regular assessments. In addition, private sector advisory committees or delegates (permanent or ad hoc/per mission) within an agency would greatly benefit the ability to properly assess status of projects, needs, recommendations from the private entity. Committees and delegates may also serve to advise on potential legal protectionary standards, and to uphold high standards of transparent communication between agency, government, international partners, and private entity.

Terra Nullius

The language used by governments and their respective agencies regarding a state presence in space must also reflect true terra nullius law. Rather than protecting state interests in space, space is to be protected by all states in collaboration for peaceful exploration and scientific advancement. Like true terra nullius law, however, states must also be ready to exercise the full extent of their jurisdiction on their space items, including debris and debris cleanup technologies and methods. In this light, states and their agencies must also adopt a more strict practice of mission review to adhere to the assessment of a mission launch or other space activities. This takes the example of requesting consultation by other international space agencies, or ethical committees/liaisons.

International Custom and Norms

Like most jurisdiction reached by international courts on the matters of high seas, terra nullius, and resource use, the laws created regarding these matters are largely based off of established norms and practices from individuals (PIs), entities (agencies), and governments, including bilateral treaties and agreements. Establishing clear good faith relations, both in practice and in treaty, will set the international framework needed to create strict and preventative protectionary, ethical, and anti-colonization methods for all future science missions. In this light, all levels of every science mission are encouraged to actively participate in effectively creating the norm of checks and balances so as to not abuse the presence of different entities, both private and public, in space. These checks may include but are not limited to: voluntarily establishing regular mandatory assessments by PPOs throughout the entirety of a mission starting from concept as well as regular mandatory assessments from international law specialists, clear and transparent international/industry partner lines of communication and delegation of authority if applicable, establishing a clear chain of command and representatives from agencies working with private industries to ensure and enforce progressive safety procedures on the private entity, strict committee/delegate presence within the private industry to communicate needs to and from the agency, and any other behavioral suggestion laid out in this paper.

Because space exploration is an extremely large, global, and mostly unified effort, governments will be at the helm of exploration and ensuring updated and proactive exploration policies for the foreseeable future. In light of this, policies and customs set to be implemented at the international level, coupled with high transparency and liaisons/infrastructure via active regional authoritative bodies in place for communicating new standards and concerns, are believed to be successful. As the world progresses in technological and scientific advancement, policy and safety standards must also remain transparent and adequately communicated across all agencies and governments. In this case, it would benefit states party to the OST to have a diplomatic office based in their country that would report directly to the UNOOSA Headquarters with matters regarding updates in biosecurity, exploration ethics, and norms and customs. Active and immediate reports to the UN from states party to the OST would strengthen international custom regarding the issues expressed in this paper. In addition, any possible infringement on UNOOSA/OST standards, or any critique raised by a space-faring nation in response to another exploration effort, would be able to be brought to the attention of the deploying nation, essentially acting as an international system of checks and balances to ensure that the priorities stated in international space law are abided by on all levels.

Utilizing the Goals of the Astro-Sciences in Collaboration
There are countless proposed and in-development instruments and missions intended to challenge our understanding of how life may exist on other worlds. The planetary protection strategy for these missions would benefit from committees made up of representatives from all potential mission stakeholders: planetary protection officers, mission astrobiologists, scientific investigators, and commercial engineers and executives. If people at every level of the mission proposal and execution process actively execute the suggestions laid out in this paper with facilitation by committees, updated and strictly heeded standards with regard to ethics, planetary protection, and communication at every phase of the mission would be much more successful and effectively become custom for future missions. These committees would benefit from planetary protection and astrobiological research well within our current capabilities, such as investigations of closed-system microbial communities on closed facilities (e.g., the ISS) and advancements in detection of organic molecules. Mission safety can also be improved by research on potentially hazardous environmental concerns, such as quantifying and mitigating radiation exposure during long-period spaceflight, the effects of planetary regolith on mechanical systems and human health, potential in-situ equipment failures, and more. Given the recent Artemis and Moon-to-Mars program announcements, as well as NASA's selection of the next New Frontiers mission, Dragonfly, these concerns only grow more time-critical.

However, such collaboration between members of the space science community and the private sector aren't limited to future planetary missions. Current commercial activity could also benefit from interdisciplinary advisory and regulatory action. The recent launch of SpaceX's Starlink internet satellite constellation raised concerns among the astronomy community due to the effects of a large artificial orbiting constellation on astronomical observing. The lack of a proactive effort to communicate with the astronomy community, is detrimental to the future of a proactive space policy and open cooperation between public and private efforts. The public outcry could have been avoided with an ethical/communicative oversight body working from mission conception to not only mitigate the constellation's effects on observing efforts, but to assess the impact of Starlink on all stakeholders. In this case, cooperative bodies made up of representatives from government agencies, academic institutions, and commercial interests can be empowered to provide effective and regular reviews of proposed commercial space activities so that any potential impacts to scientific activity (or the integrity of other ground- and space-based environments) can be minimized. This would allow for innovation and competition from the private sector to continue to flourish while holding private science to the standards used by government agencies and public science. This recognizes that public science adheres to strict safety and ethical practice, and "red tape" in the context of privatized ventures into space is necessary. Private industry must also recognize the necessity of transparent technological standards and technology sharing[10], as programs like Starlink and other space-based communications infrastructure. These technologies put forth by the private industry with global needs in mind, though they have the ability to benefit communities worldwide, must also recognize that in the mind of one private entity, "global good" may not be an inherent "good" in the eyes of all communities worldwide. Thus, global efforts would have to go through global vetting. With these concerns in mind, we reiterate that commercial and scientific activity must avoid harm to both terrestrial and space environments, with particular emphasis on maintaining pristine environments and Native lands.[10]

## The Suggested Framework

Suggestions laid out in this paper regarding ethical implications, safety assurances, contamination control and planetary protection, communication, and accountability standards would ideally fall within the earliest stages of the science traceability matrix. Continued and mandatory assessments of these standards throughout the mission must also be coupled with early traceability to ensure that these ideals remain throughout entire mission timeline, and throughout all aspects of the mission. This way, a consistent watch on ethics, communication, planetary protection, and other policy and social concerns via committees and liaisons catches any possible infringements, and pushes the international communities to adopt more updated and proactive standards. Running parallel to internal and transparent monitoring of ethics and communication is the internationally communicative component, capable of acting as a "check" in international relations, and turning reactive international policy into proactive and constantly-updated policies that can be used by a variety of agencies and nations. Though this paper uses the example of NASA and the United States Government, the framework of transparent and proactive standards that are open to critique and custom-setting can be applied to all governments, institutions, and missions.

# References.

[1] National Academies of Sciences E. *Review and Assessment of Planetary Protection Policy Development Processes.*; 2018. doi:10.17226/25172

[2] Office of Safety and Mission Assurance. *Biological Contamination Control for Outbound and Inbound Planetary Spacecraft (NPD 8020.7G).* NASA Policy Directive February 19, 1999.

[3] United Nations Office for Outer Space Affairs. *Treaty on Principles Governing the Activities of States in the Exploration and Use of Outer Space, including the Moon and Other Celestial Bodies.* The United Nations December 19, 1966.

[4] Office of Safety and Mission Assurance. *Planetary Protection Provisions for Robotic Extraterrestrial Missions (NPD 8020.12D).* NASA Policy Directive April 20, 2011.

[5] Journal of the British Interplanetary Society, Volume 67, No 1, pp 1-43.

[6] Vakoch, D. A. (2016). In defence of METI. Nature Physics, 12(10), 890.

[7] United Nations Office for Outer Space Affairs. *Convention on Registration of Objects Launched into Outer Space (3235 XXIX) Resolution adopted by the General Assembly.* The United Nations November 12, 1974.

[8] Knapp, Alex. "Understanding The Thirty Meter Telescope Controversy." *Forbes*, Forbes Magazine, 12 June 2015.

[9] "Legal Process – Thirty Meter Telescope." *Mauna Kea and TMT*, Thirty Meter Telescope.

[10] United Nations Educational, Scientific, and Cultural Organization. "Records of the General Conference, 33rd Session, Paris 3-21 October 2005 V.1: Resolutions ." *Universal Declaration on Bioethics and Human Rights, Article 18*